\documentclass[prb,showpacs,footinbib,twocolumn,final,superscriptaddress]{revtex4-1}

\usepackage{pslatex}
\usepackage{amsfonts,amsmath,amssymb,ifpdf,epsfig,graphicx}
\usepackage[usenames,dvipsnames]{color}
\usepackage{hyperref}
\usepackage{bm}
\usepackage{comment}

\newcommand{\vq}{{\bf{q}}}
\newcommand{\vQ}{{\bf{Q}}}
\newcommand{\vQa}{{\bf{Q_1}}}
\newcommand{\vQb}{{\bf{Q_2}}}
\newcommand{\vk}{{\bf{k}}}

\newcommand{\vx}{{\bf{x}}}
\newcommand{\vy}{{\bf{y}}}

\newcommand{\vd}{{\bf{d}}}
\newcommand{\vE}{{\bf{E}}}
\newcommand{\vN}{{\bf{0}}}

\newcommand{\wo}{\omega}
\newcommand{\ep}{\epsilon}
\newcommand{\eF}{{\epsilon_{\rm F}}}
\newcommand{\eFtyp}{\mu^{\,0}}
\newcommand{\vF}{{v_{\rm F}}}

\newcommand{\ind}{n_{\rm ind}}

\newcommand{\ETh}{E_{\rm Th}}
\newcommand{\tPhi}{\tau_\varphi}
\newcommand{\gammaZD}{\gamma_{\rm 0D}}
\newcommand{\gammaErg}{\gamma_{\rm erg}}
\newcommand{\tDw}{\tau_{\rm dw}}
\newcommand{\Tcross}{T_{\rm cross}}
\newcommand{\Tsat}{T_{\rm sat}}

\newcommand{\dx}[1]{ \! {\rm d}#1 \, }
\newcommand{\dcx}[1]{ \! {\rm d}^3#1 \, }

\newcommand{\Eq}[1]{Eq.~\eqref{#1}}
\newcommand{\EqPlain}[1]{\eqref{#1}}
\newcommand{\EqsFromTo}[2]{Eqs.~(\ref{#1})-(\ref{#2})}

\newcommand{\EqsTwo}[2]{Eqs.~(\ref{#1},\ref{#2})}
\newcommand{\EqsThree}[3]{Eqs.~(\ref{#1}), (\ref{#2}) and (\ref{#3})}
\newcommand{\Fig}[1]{Fig.~\ref{#1}}
\newcommand{\Figs}[1]{Figs.~\ref{#1}}
\newcommand{\FigsTwo}[2]{Fig.~\ref{#1} and \ref{#2}}
\newcommand{\Sect}[1]{Section~\ref{#1}}

\newcommand{\Ref}[1]{Ref.~[\onlinecite{#1}]}
\newcommand{\RefsTwo}[2]{Ref.~[\onlinecite{#1}] and [\onlinecite{#2}]}

\begin{document}

\title{Quantum Corrections to the Polarizability and Dephasing in Isolated Disordered Metals}

\author{M.~Treiber}
\affiliation{Ludwig Maximilians University, Arnold Sommerfeld Center
             and Center for Nano-Science, Munich, D-80333, Germany}


\author{P.~M.~Ostrovsky}
\affiliation{Max Planck Institute for Solid State Research, Heisenbergstr. 1, 70569 Stuttgart, Germany}
\affiliation{L. D. Landau Institute for Theoretical Physics, 142432 Chernogolovka, Russia}

\author{O.~M.~Yevtushenko}
\affiliation{Ludwig Maximilians University, Arnold Sommerfeld Center
             and Center for Nano-Science, Munich, D-80333, Germany}

\author{J.~von~Delft}
\affiliation{Ludwig Maximilians University, Arnold Sommerfeld Center
             and Center for Nano-Science, Munich, D-80333, Germany}

\author{I.~V.~Lerner}
\affiliation{School of Physics and Astronomy, University of Birmingham,
             Birmingham, B15 2TT, UK}

\date{\today}

\begin{abstract}
  We study the quantum corrections to the polarizability of isolated
  metallic mesoscopic systems using the loop-expansion in diffusive
  propagators. We show that the difference between connected
  (grand-canonical ensemble) and isolated (canonical ensemble) systems
  appears only in subleading terms of the expansion, and can be
  neglected if the frequency of the external field, $\, \wo $, is of the
  order of (or even slightly smaller than) the mean level spacing,
  $ \, \Delta $. If $ \, \wo \ll  \Delta $, the two-loop correction
  becomes important. We calculate it by systematically evaluating the
  ballistic parts (the Hikami boxes) of the corresponding diagrams and
  exploiting electroneutrality. Our theory allows one to
  take into account a finite dephasing rate, $ \, \gamma $, generated
  by electron interactions, and it is complementary to the
  nonperturbative results obtained from a combination of random matrix
  theory (RMT) and the $ \, \sigma $-model, valid at $\gamma \to 0$.
  Remarkably, we find that the two-loop result for isolated systems with
  moderately weak dephasing, $\gamma \sim \Delta$, is similar to the
  result of the RMT+$\sigma$-model even in the limit $\wo \to 0$.
  For smaller $\gamma$, we discuss the possibility to interpolate
  between the perturbative and the nonperturbative results. We compare
  our results for the temperature dependence of the polarizability of
  isolated rings to the experimental data of
  R. Deblock \emph{et al.} [\prl \ {\bf 84}, 5379 (2000); \prb \ {\bf 65}, 075301 (2002)], and we
  argue that the elusive 0D regime of dephasing might have manifested
  itself in the observed magneto-oscillations. Besides, we thoroughly
  discuss possible future measurements of the polarizability, which
  could aim to reveal the existence of 0D dephasing and the role of the
  Pauli blocking at small temperatures.
\end{abstract}

\maketitle


\section{Introduction}
\label{sect:Introduction}

Interference phenomena in mesoscopic electronic systems require phase
coherence, which is cut beyond the so-called dephasing time $\, \tau_{\phi} $.
At low temperatures $T \lesssim 1K$, where phonons are frozen out, dephasing
is caused mainly by electron interactions, which lead to a finite dephasing rate
\cite{Altshuler_Disordered_1985} $ \, \gamma \equiv 1/\tau_{\phi}$.
In large systems  with a small
Thouless energy, $\ETh \ll T$, dephasing crucially depends on dimensionality
and geometry\cite{Altshuler_SmallEnergyTransfers_1982}.
However, if the system is finite and $T \lesssim \ETh$, spatial
coordinates become unimportant and a 0D regime of rather
weak dephasing is expected to occur \cite{Sivan_QuasiParticleLifetime_1994}.
This regime is characterized by a universal temperature dependence
of the dephasing rate, $\gammaZD \sim \Delta T^2 / \ETh^2$, where $ \,
\Delta \, $ is the mean-level spacing.
This $T$-dependence of $\gamma$ can be explained by simple power counting:
Pauli blocking restricts  the number of available final scattering
states of the electrons, therefore both the energy transfer and the
available phase-space are $ \, \propto T$, similar to the standard result
for a clean Fermi-liquid.
However, despite the fundamental nature and  the physical importance of
0D dephasing, attempts to observe it experimentally in mesoscopic systems
have been unsuccessful so far.

In transport experiments, the 0D regime is generally difficult to
observe, since quantum transport is almost insensitive to $\gamma$
at $T \ll \ETh$. For example, the weak localization correction to
the  classical dc conductivity
is cut mainly by the dwell time, $ \, \tDw \ll  1/\gammaZD $,
see  \Ref{Treiber_QuantumDot_2012} for a detailed discussion.
This is an unavoidable problem which occurs in any open system even if the
coupling to leads is weak.

In this work, we concentrate on interference phenomena in isolated systems,
where $ \, \tDw \to \infty \, $ and where 0D dephasing is not masked by the
coupling to the environment. Deeply in the 0D regime at $ \, \gamma \ll \Delta \, $,
the spectrum of the isolated system is discrete
\cite{Altshuler_QuasiParticleLifetime_1997,Blanter_ScatteringRate_1996}
and, in the absence of other mechanisms of dephasing, random matrix theory
(RMT) can be used as a starting point for an effective low-energy theory
at $ \, E\ll \ETh $ \cite{Kravtsov_Corrections_1994,Blanter_Polarizability_1998}.
Unfortunately, RMT is not appropriate for a systematic account of dephasing.

If one is interested in the (almost 0D) regime $ \, \gamma \le \Delta $, where
the spectrum is not yet discrete, the usual mesoscopic perturbation theory
\cite{Montambaux_Book_2007} can be used, which is able to take into account
dephasing in all regimes. However, the description of quantum effects in
isolated systems provides a further technical challenge. Namely, the usual
perturbation theory is well developed for a fixed chemical potential $\mu$;
i.e. it describes systems in the grand-canonical ensemble (GCE). Realizing
the canonical ensemble (CE), where the number of particles is fixed instead,
can be rather tricky, see, e.g., \Ref{Kamenev_StatEnsemble_1993}.
In the following, we assume that a description in terms of the so-called
Fermi-level pinning ensemble introduced in
\RefsTwo{Lehle_Canonical_1995}{Altland_Canonical_1992}
is applicable \cite{remark_canonical}.

The dephasing rate of an isolated mesoscopic system can be explored, for
instance, by measuring quantum components of the electrical polarizability
$\alpha$ at a given frequency $ \, \wo$:
\begin{equation}\label{eq:defAlpha}
   \alpha(\wo) = \vd(\wo) \cdot \vE(\wo)/|\vE(\wo)|^2 \, .
\end{equation}
Here $ \, \vE \, $ is a spatially homogeneous electric field and $ \, \vd \, $
is the total induced dipole moment in the sample.

Gorkov and Eliashberg studied the polarizability in the
seminal work \Ref{Gorkov_Eliashberg_1965} by using
results from RMT and found very large quantum corrections. Later, it
was shown in \Ref{Rice_Polarizability_1973} that the corrections are
significantly reduced if screening is taken into account
correctly \cite{Blanter_GorkovEliashberg_1996}.
Efetov reconsidered Gorkov and Eliashberg's calculation in
\Ref{Efetov_Polarizability_1996} and derived a formula which accounts
for screening in the random phase approximation (RPA) and expresses the
quantum corrections to $\alpha$ in terms of correlation functions of
the wave-functions and energy levels of the system.
Noat \emph{et al.} \cite{Noat_Polarizability_1996} used a simple model
supported by numerical simulations to  analyze the difference between the GCE and the CE,
and established that the quantum corrections are always small for systems
with a large dimensionless conductance.
Subsequently,
Mirlin and Blanter \cite{Blanter_Polarizability_1998} studied the polarizability
using a combination of RMT and the diffusive $\sigma$-model. In particular, they have calculated
$\wo$-dependence of $\alpha$ at $\wo \ll \ETh$ for the case of the CE at $ \, \gamma = 0 $.
Thus, neither the temperature nor the magnetic field dependence of $ \, \alpha \, $
has been described until now.

Besides the progress made in theory, experimental measurements
of the quantum corrections have been reported in
\RefsTwo{Deblock_Experiment_2000}{Deblock_Experiment_2002}.
The authors  measured the $ \, T$-dependence of the polarizability of
small metallic rings placed in a superconducting resonator (with a fixed
frequency $\wo$) in a perpendicular magnetic field and tried to extract
the $ \, T$-dependence of $ \, \tau_{\phi} \, $ by using an empirical
fitting equation. A fingerprint of 0D dephasing was found at low
temperatures, though a reliable identification of the temperature dependence
of $\tPhi$ calls for a more rigorous theory.

Motivated by the experimental results, we develop a perturbative
theory for the quantum corrections $\Delta \alpha$ to the polarizability by
using the mesoscopic ``loop-expansion'' in diffusons and Cooperons, where
$ \, \gamma \, $ plays the role of a Cooperon mass. We have chosen the
experimentally relevant parameter range $ \, \max(\wo,\gamma) \gtrsim \Delta \, $.
Generically, the difference between the GCE and the CE can be  important
up to energies substantially exceeding $ \, \Delta $, see the discussions
in \Ref{Kamenev_StatEnsemble_1993}.
To check whether this statement also applies for $ \, \Delta \alpha $,
we calculate leading and subleading corrections in the Fermi-level
pinning ensemble. The former corresponds solely to the one-loop
answer of the GCE while the latter includes the two-loop answer of the GCE
and additional terms generated by fixing the number of particles in the CE.
We show that, within our approach, the leading term of the perturbative
expansion for $ \,\Delta \alpha \, $
suffices for its theoretical description in the experimentally relevant
parameter range of \RefsTwo{Deblock_Experiment_2000}{Deblock_Experiment_2002}.
This important result of the present paper allows us to find the dependence of
$ \, \Delta \alpha \, $ on temperature and on magnetic field.
Our theoretical results are in good qualitative agreement with the experiments,
though we show that the present experimental data are not sufficient
for a reliable identification of 0D dephasing. We suggest repeating the
experimental measurements with higher precision and lower frequencies and
using the fitting procedures which  we propose in the present paper.
We have good hopes that the elusive 0D regime of dephasing
may be detectable in this manner in the near future.

The rest of this paper is organized as follows:

\vspace{2mm}
\Sect{sect:Polarizability}: 
we derive a general expression for the polarizabiliy as a functional of the density response function in the RPA.

\vspace{2mm}
\Sect{sect:DensityResponseFunction}: 
we calculate the leading quantum corrections of the density response
function for connected as well as isolated disordered metals.
This part of the paper is rather formal and
technical. Readers who are not interested in details of
the calculations can safely skip it, paying attention only to our key results, which we list here.
First, we derive the
one- and two-loop quantum corrections for the GCE which are presented in \EqsTwo{eq:GCE1}{eq:GCE2}
of Subsection \ref{sect:DensityResponseFunction-GCE}. A ``naive'' loop-expansion for the
GCE suffers from a double-counting problem of some diagrams which leads to a violation
of the particle  conservation law (electroneutrality) accompanied by artificial UV divergences.
We suggest an algorithm of constructing the diagrams which allows one to avoid all these problems.
Our method can be straightforwardly checked for the one-loop calculations, see \Fig{fig:Hikami},
and we extend it to the much more cumbersome two-loop diagrams shown in \Fig{fig:GCE2}.
Second, we calculate the leading diagrams which appear due to fixing the Fermi level in the CE.
Their contribution is given by \Eq{eq:CE1} of Subsection~\ref{sect:DensityResponseFunction-CE}.

\vspace{2mm}
\Sect{sect:QuantumCorrections}: 
we use the results from \Sect{sect:DensityResponseFunction} to derive
a general equation for the quantum corrections $\Delta \alpha$.

\vspace{2mm}
\Sect{sect:ComparisonRMTSigma}: 
we compare our findings to the results obtained from a combination of the RMT and the
$\sigma$-model. We show that the diagrammatic result in the limit of a large conductance, \Eq{eq:DeltaFSimple},
qualitatively reproduces all features of the nonperturbative answers for almost 0D systems
at $ \, 0 \le \omega < \ETh $, see \Fig{fig:RMT}.

\vspace{2mm}
\Sect{sect:ComparisonExperiment}: 
we apply our results for $\Delta
\alpha$ to the ring geometry, present a comparison with previous experiments
and discuss possible future measurements which can reliably
confirm the existence of 0D dephasing.

\section{Polarizability}
\label{sect:Polarizability}

The polarizability \EqPlain{eq:defAlpha} is governed by the
induced dipole moment in the sample,
\begin{align}\label{eq:defDipole}
  \vd(\wo)
  = \int_V \dcx{\vx} \left[ \vx \cdot \ind(\vx,\wo) \right] \, ,
\end{align}
where $\ind$ is the induced charge density.
In the case of a good metal,
screening should be taken into account in the random phase approximation
(RPA), which results in the following expressions for the Fourier
transform of $\ind$:\cite{Bruus_Book_2004}
\begin{align}\label{eq:defInd}
  \ind(\vq,\wo) =
  -2 e^2 \frac{\chi(\vq,\wo)}{\ep(\vq,\wo)} \phi_{\rm ext}(\vq,\wo) \, .
\end{align}
Here  $\phi_{\rm ext}(\vx,\wo)  = - \vE(\wo) \cdot \vx$ is the external
electric potential, $\ep(\vq,\wo) = 1 - 2U(\vq)\chi(\vq,\wo)$ is the
dielectric function, $U$ is the bare Coulomb potential, and  $\chi$ is the
density response function per spin. By using the Kubo formula, $\chi$
can be expressed in terms of the commutator of the density operators $ \, \hat{n} $:
\begin{align}\label{eq:Kubo}
  \chi(\vq,\wo) =  i \int_V \dcx{\vx} \ \int_0^{\infty} \!\!\! \dx{t} \, \langle \left[ \hat{n}(\vx,t),
       \hat{n}(\vN,0) \right] \rangle e^{-i ( \vq \vx - \wo t ) } \, .
\end{align}
We assume spatial homogeneity of the system, which is restored after disorder averaging.

Inserting \EqsTwo{eq:defDipole}{eq:defInd} in \Eq{eq:defAlpha}, we find
the following expression for the polarizability:
\begin{equation}\label{eq:AlphaInitital}
  \alpha(\wo) \!=\! \frac{2 e^2}{|\vE(\wo)|^2}
  \frac{1}{V} \sum_{\vq \neq \vN} \ \phi_{\rm ext}(\vq,\wo)
  \frac{\chi(\vq,\wo)}{\ep(\vq,\wo)} \phi_{\rm ext}(-\vq,\wo) \
   \, .
\end{equation}
Note that the zero-mode does not contribute to $\alpha$ because of
{\it electroneutrality of the sample}:
\begin{equation}\label{eq:Electroneutrality}
  \chi(\vq \equiv \vN,\wo) = 0 \, .
\end{equation}

For a clean metal at $ \wo \ll \vF \vq $ ($ \vF \, $ is the Fermi velocity),
$\chi$ is
local and is given by the density of states at the Fermi level:
\begin{align}\label{eq:ChiClassical}
  \chi(\vq,\wo \to 0) = \rho_0 \, .
\end{align}
The same equation
holds true for a disordered (classical) metal at $ \wo \ll D \vq^2 $
($ D \, $ is the diffusion constant), see \Sect{sect:DensityResponseFunction}.
Eqs.(\ref{eq:AlphaInitital},\ref{eq:ChiClassical})
yield the ''classical`` polarizability $\alpha_0$ of the disordered sample.

\section{Density response function}
\label{sect:DensityResponseFunction}

In this section, we consider the density response function of the
disordered metal which is needed to calculate the polarizability,
\Eq{eq:AlphaInitital}. We will start with the loop-expansion of
the disorder-averaged $ \, \chi \, $ in the GCE: $ \, \overline{\chi}|_{\mu = {\rm const}}
\equiv \overline{\chi}_{\mu } $. It is relevant for the polarizability of
the connected system. Besides, the two-loop contribution to
$ \, \overline{\chi}_{\mu} \, $ is needed to study the difference between
the answers in the GCE and the CE. The latter is described in the second
part of the present section.

We consider only weakly-interacting disordered systems at small
temperatures. The main role of the electron interaction is to generate
a finite $ \, T $-dependent dephasing rate for Cooperons. Therefore,
we derive the density response function for the non-interacting system
at $ \, T = 0 \, $ and take into account $ \, \gamma(T) \, $ at the end
of the calculations.

\subsection{Grand canonical ensemble}
\label{sect:DensityResponseFunction-GCE}

Simplifying \Eq{eq:Kubo} for the non-interacting system at $ T = 0 $ and
fixed $ \, \mu$, $ \, \chi_{\mu} \, $ can be
presented in terms of retarded/advanced ($ G^{R/A} $) Green's functions
(GFs) \cite{Montambaux_Book_2007}:
\begin{align}\label{eq:chiMu}
   & \chi_{\mu}(\vx,\vy,\wo)
  = - \int_{-\infty}^{0} \dx{\ep} \\\nonumber
  & \times
    \bigg( \rho_{\mu+\ep}(\vx,\vy) G^A_{\mu+\ep-\wo}(\vy,\vx)
      + G^R_{\mu+\ep+\wo}(\vx,\vy) \rho_{\mu+\ep}(\vy,\vx) \bigg) \, .
\end{align}
Here we have introduced the spectral function
(or the non-local density of states):
\begin{align}
\label{eq:defRho}
  \rho_\ep(\vx,\vy)
  & \equiv \frac{i}{2\pi}
           \left[ G_\ep^{R}(\vx,\vy) - G_\ep^{A}(\vx,\vy) \right] \, .
\end{align}

\begin{figure}[t]
  \begin{center}
	\includegraphics[width=0.99\columnwidth]{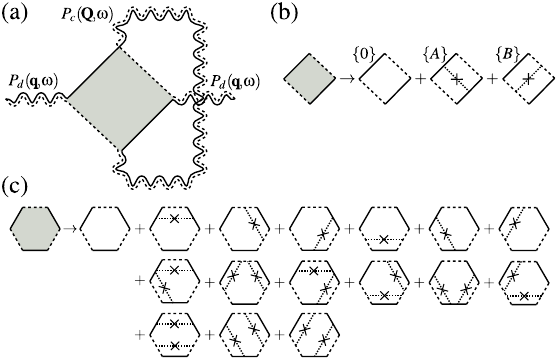}
  \end{center}
	\caption{(a): One-loop correction to the density response function in
	         the GCE. Retarded (advanced) GFs are denoted by solid (dashed)
	         lines. Impurity lines, corresponding to the correlation function
	         \EqPlain{eq:WhiteNoise}, are denoted by dotted crossed lines.
           Diffusive propagators are represented by wavy double lines.
           They denote impurity ladders between the corresponding GFs of
           opposite retardation either in the particle-particle (Cooperon,
           $P_c$) or in the particle-hole (diffuson, $P_d$)
           channel. (b) and (c): Dressed 4- and 6-point Hikami boxes which include
           diagrams with one or two additional impurity lines connecting GFs of the
           same retardation.
          }
	\label{fig:GCE1}
\end{figure}

In the presence of a random Gaussian white-noise disorder potential $V(\vx)$
with correlation function
\begin{align}\label{eq:WhiteNoise}
   \overline{ V(\vx)V(\vy) } = \frac{1}{2\pi \rho_0 \tau} \delta(\vx-\vy) \, ,
\end{align}
the disorder-averaged GFs are given by
\begin{align}\label{eq:AveragedGFs}
   \overline{G}^{R/A}_{\ep}(\vk) = \frac{1}{\ep - \ep_\vk \pm i/2\tau},
\end{align}
where $\tau$ is the impurity scattering time and $ \, \ep_\vk \, $ is
the particle dispersion relation.

The disorder average of \Eq{eq:chiMu} can be calculated with the help
of the usual diagrammatic methods,\cite{Altshuler_Disordered_1985}
which yield the loop-expansion:
\begin{align}
   \overline{\chi}_{\mu}(\vq,\wo)
   = \chi_0(\vq,\wo) + \sum_{j} \delta \chi_{\rm GCE}^{(j)} \, .
\end{align}
Here $ \, j \, $ is the number of loops built from impurity ladder
diagrams which include ladders in the particle-hole channel (diffuson
propagators) or in the particle-particle channel (Cooperon propagators).
The leading (classical) term is well-known\cite{Altshuler_Disordered_1985}:
\begin{align}\label{eq:chi0}
   \chi_0(\vq,\wo) = &\ \rho_0 \frac{D\vq^2}{D\vq^2 - i\wo} \, .
\end{align}
It obeys the fundamental requirement of electroneutrality, \Eq{eq:Electroneutrality},
and reduces to \Eq{eq:ChiClassical} at $\wo \ll D\vq^2$.

The leading quantum correction $ \, \delta \chi_{\rm GCE}^{(1)} \, $
describes the weak-localization correction to the diffusion constant
\cite{Vollhardt_Localization_1980} and, therefore, is also well-known.
Nevertheless, we would like to recall the basic steps of its derivation, which
will be important to find the more complicated subleading term
$ \delta \chi_{\rm GCE}^{(2)} $.

The one-loop diagram, which yields  $ \, \delta \chi_{\rm GCE}^{(1)} $,
is shown  in \Fig{fig:GCE1}(a). It includes two diffuson propagators
$P_d$ and one Cooperon propagator $P_c$, which
are given by
\begin{align}\label{eq:Diffuson}
  P_d(\vq,\wo) = \frac{1}{D\vq^2 - i\wo} \, ,
  \quad
  P_c(\vQ,\wo) = \frac{1}{D\vQ^2 - i\wo + \gamma} \, .
\end{align}
The (ballistic) part of
the diagram which connects the diffusive propagators is known
as a 4-point Hikami box \cite{Hikami_AndersonLocalization_1981}.
It consists of three diagrams of the same order in $(\eF \tau)^{-1}$
shown in \Fig{fig:GCE1}(b) and labeled by $\{0\}$, $\{A\}$, and $\{B\}$, which are
obtained by inserting additional impurity lines between GFs of the same
retardation (``dressing'' the  Hikami box). The
Hikami box should be calculated by expanding the GFs in each of the three
diagrams in the transferred momenta and energies.
A direct summation of the three diagrams gives
\begin{align}\label{eq:H4DirectSum}
  H_4^{\rm (direct \ sum)} = 4\pi\rho_0 \tau^4 \left[ D\vq^2 + D\vQ^2 - i\wo \right] \, .
\end{align}
The second and third terms in parentheses are manifestly incorrect as they
violate electroneutrality, \Eq{eq:Electroneutrality}, and lead to an unphysical
UV divergence in  $ \, \delta \chi_{\rm GCE}^{(1)} $. The incorrect terms originate
from a double-counting problem: the diagram with a single impurity line, which
contributes (via the diffuson) to the classical result of \Eq{eq:chi0},
is also included in the quantum correction  $ \, \delta \chi_{\rm GCE}^{(1)} $
via the Cooperon attached to the ``undressed'' part of the Hikami box -- the empty
square $\{0\}$. One can eliminate unphysical UV divergent diagrams in the framework 
of the nonlinear $\sigma$-model by choosing an appropriate parametrization of
the matrix field \cite{Ef:83,pavel_unpublished}. However, to the best of our knowledge a
consistent procedure of their elimination in the framework of straightforward
diagram techniques was not described in literature. As this is rather important
for any calculation beyond the one-loop order, we give a detailed description of
such a procedure below.

\begin{figure}[t]
  \begin{center}
	\includegraphics[width=0.99\columnwidth]{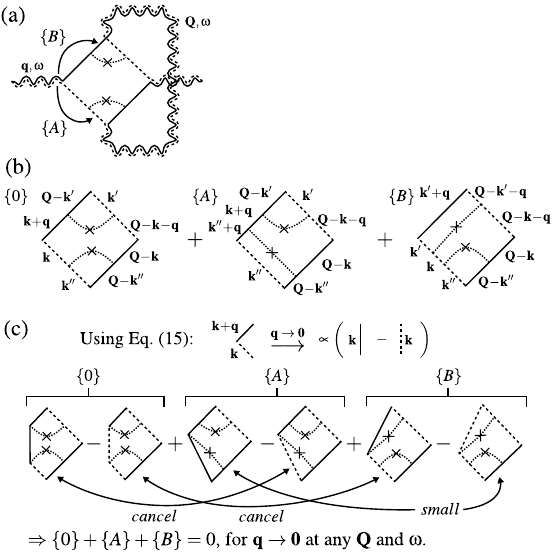}
  \end{center}
	\caption{(a) The ``skeleton diagram'', which we use to generate the
	dressings $\{A\}$ and $\{B\}$ of the Hikami box shown in \Fig{fig:GCE1}(b).
	The arrows with labels $\{A\}$ and $\{B\}$ indicate how the (diffuson attached) external
                vertex has to be moved to generate the correponding dressed boxes.
	(b) The resulting diagrams with the undressed, $\{0\}$, and two dressed boxed can be
                summed up directly, since no double-counting
	problem appears. To leading order in the transferred momenta and energies, $(D\vq^2\tau,D\vQ^2\tau,\wo\tau) \ll 1$,
	the sum of the three diagrams in (b) is $4\pi\rho\tau^4 D \vq^2$.
	(c) Dressing the Hikami box by moving the external vertex guarantees that
	the answer vanishes at $\vq \to \vN$, since the 3 
	diagrams either cancel each other exactly (at any $\vQ$ and $\wo$), or are small in this limit.
	This can be seen immediately after using the identity \EqPlain{eq:GFRAIdentity} and redrawing the
                boxes $\{0\}$,  $\{A\}$ and $\{B\}$
	as the 6 diagrams shown in the last line.}
	\label{fig:Hikami}
\end{figure}

To avoid the double-counting, the Cooperon ladder of \Fig{fig:GCE1}(a)
should start with \emph{two}
impurity lines when attached to the undressed box, while it should still start
with one impurity line when attached to the dressed box. Thus, there is an ambiguity
in the independent definition of the Hikami boxes and the ladder diagrams.
We suggest a general algorithm which allows us to overcome this
ambiguity and generate all properly dressed Hikami boxes obeying
electroneutrality\cite{Hastings_Inequivalence_1994,pavel_unpublished}. 

Let us consider the 4-point Hikami box shown in \Fig{fig:Hikami}(a) to illustrate the method. 
\Fig{fig:Hikami}(a) is obtained from \Fig{fig:GCE1}(a) by
``borrowing'' two impurity lines to the undressed Hikami box from the attached Cooperon.
We use this undressed box in \Fig{fig:Hikami}(a)
as a ``skeleton diagram'' which generates the dressings $\{A\}$ and $\{B\}$ of \Fig{fig:GCE1}(b)
by moving one of the  external vertices (with diffuson attached) past one
of the borrowed impurity lines. Two possible movements of the left
external vertex are indicated by arrows with labels $\{A\}$ and
$\{B\}$ in \Fig{fig:Hikami}(a). \Fig{fig:Hikami}(b) shows all three
components of the fully dressed Hikami box: two generated
boxes, $\{A\}$ and $\{B\}$, and the undressed box, $\{0\}$, where
the external vertex is not moved.
Dressing the Hikami box in this way removes the
ambiguity, since  all the Cooperon  ladders attached to each of the boxes start with two impurity lines, thus avoiding the
double-counting.
Furthermore, using the identity\cite{Hastings_Inequivalence_1994}
\begin{align}\label{eq:GFRAIdentity}
  \overline{G}^R_{\ep+\wo}(\vk\!+\!\vq)\overline{G}^A_\ep(\vk) \ \overset{\vq \to 0}{\longrightarrow} \
  \frac{i\tau}{1-i\tau\wo} \left[\overline{G}^R_{\ep+\wo}(\vk) - \overline{G}^A_{\ep}(\vk)\right] \, ,
\end{align}
we illustrate in \Fig{fig:Hikami}(c) that  in the limit $\vq \to 0$  the
generated diagrams automatically cancel each other [to leading order in
$ \, (\eF \tau)^{-1} \ll 1 \, $] at any $\vQ$ and $\wo$,
thus ensuring electroneutrality and the absence of the UV divergence.

Summing up the 3 diagrams drawn in \Fig{fig:Hikami}(b)
and using the resulting expression to calculate the diagram
shown in \Fig{fig:GCE1}(a), we obtain the well-known result\cite{Vollhardt_Localization_1980}
\begin{align}\label{eq:GCE1}
   \delta \chi_{\rm GCE}^{(1)}(\vq,\wo)
   = & \ \frac{1}{\pi V} \frac{D\vq^2 i\wo}{(D\vq^2 - i\wo)^2}
         \sum_\vQ P_c(\vQ,\wo) \, .
\end{align}
Note that $ \, \delta \chi_{\rm GCE}^{(1)} / \chi_0 \sim {\cal O}( \Delta / \max(\wo,\gamma)) $,
where $\Delta \equiv 1/(\rho_0 V)$.
Thus, \Eq{eq:GCE1} describes the dominating quantum correction to $ \,
\overline{\chi}_\mu \, $ if $\max(\wo,\gamma) \gg \Delta $.

\begin{figure}[t]
  \begin{center}
	\includegraphics[width=0.99\columnwidth]{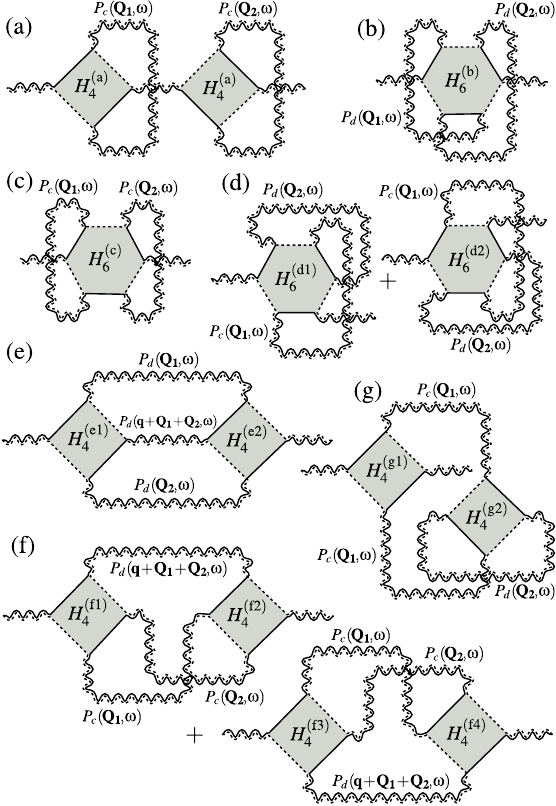}
  \end{center}
	\caption{Diagrams contributing to the two-loop correction
	         $ \, \chi_{\rm GCE}^{(2)} $. Answers for the Hikami boxes read:
           $H_4^{\rm (a,g1)} =   {\cal D} \vq^2$, \
	         $H_6^{\rm (b,c)} = - \tau^2 {\cal D} \vq^2$, \
	         $H_6^{\rm (d1,d2)} = 0$, \ $H_4^{\rm (e1)}\times H_4^{\rm (e2)} =  ({\cal D} (\vq(\vq+\vQa+\vQb)))^2$, \
	         $H_4^{\rm (f1)}\times H_4^{\rm (f2)} = H_4^{\rm (f3)}\times H_4^{\rm (f4)} = 2 {\cal D}^2 (\vq\vQa)(\vq\vQb)$, and
	         $H_4^{\rm (g2)} =  {\cal D}( \vQa^2 + \gamma/D )$,
	         see the main text for details. Here $ \, {\cal D} = 4\pi\rho\tau^4 D$.
          }
	\label{fig:GCE2}
\end{figure}
\begin{figure}[t]
  \begin{center}
	\includegraphics[width=0.99\columnwidth]{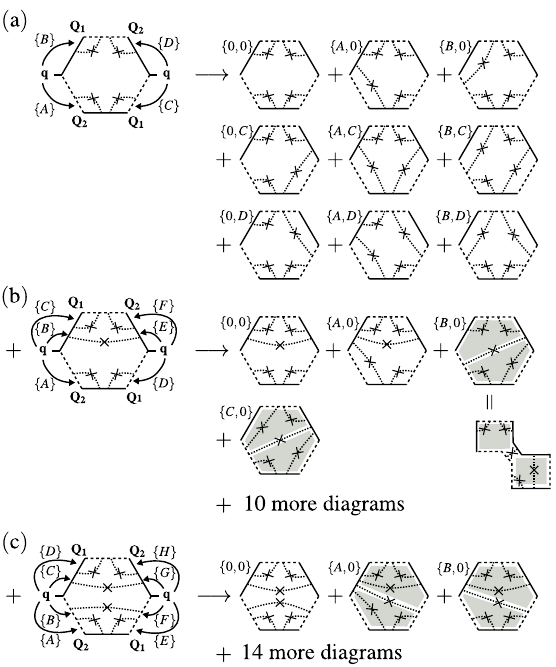}
  \end{center}
	\caption{Dressing of the 6-point Hikami box of \Fig{fig:GCE2}(b)
           using the algorithm introduced in \Fig{fig:Hikami}.
           (a) Only 8 of the 15 dressings are generated by moving the vertices.
           The other dressings are generated by adding one (b) or two (c)
           impurity lines, followed by repeating the procedure.
           This algorithm also generates products of 4-point Hikami boxes,
           indicated by a gray box.
           Summing up all 40 diagrams yields $ -12\pi\rho \tau^6 D\vq^2$.
          }
	\label{fig:Hikami6}
\end{figure}

To calculate the subleading quantum corrections, one has to consider the
two-loop diagrams shown in \Fig{fig:GCE2}, which contain momentum sums
over diffuson or Cooperon propagators, or both. Thus, their
contribution is subleading in either $ (\Delta / \max(\wo,\gamma)) $,
$ (\Delta / \wo) $, or $ (\Delta / D\vq^2) $. Note that the diagrams
containing only diffusons are not relevant for the experiments, since
they are magnetic field independent. We have used the algorithm
described above to calculate the 4-point Hikami-boxes $H_4^{(a)-(g1)}$ of \Fig{fig:GCE2} avoiding double-counting
and maintaining electroneutrality, \Eq{eq:Electroneutrality}. 
The ``inner'' Hikami box of \Fig{fig:GCE2}(g), $H_4^{(g2)}$, is of different nature because
it is connected to two internal Cooperons. Nevetheless, the same double counting problem appears
and can be overcome with the help of dressing this box by moving the vertices with the attached Cooperons.
As a result, electroneutrality does not necessarily apply for $H_4^{(g2)}$, which is reflected by its
$ \, \gamma $-dependence,  see the next paragraph.
Besides, the diagrams shown in \Fig{fig:GCE2}(b-d) contain 6-point
Hikami boxes. Their dressing is more subtle because of two issues, see
the example shown in \Fig{fig:Hikami6}, which corresponds to the Hikami
box $H_6^{\rm (b)}$ of \Fig{fig:GCE2}(b): First, starting with the undressed diagram
and moving vertices into the attached diffusons, one cannot generate all
required 15 dressings shown in \Fig{fig:GCE1}(c). Instead, only 8
dressings can be obtained for the 6-point Hikami box, cf. \Fig{fig:Hikami6}(a).
That problem can be solved by considering two more ``skeleton diagrams''
with one-,  \Fig{fig:Hikami6}(b), and two-,  \Fig{fig:Hikami6}(c),
additional impurity lines between GFs of the same retardation.
All of the missing dressings can be obtained by applying the above
described algorithm similar to \Fig{fig:Hikami6}(a). Second, by moving
the vertices of the diagrams in \Figs{fig:Hikami6}(b,c) new diagrams of
the same order in $(\eF \tau)^{-1} \ll 1$ are generated, which look like
products of two dressed or undressed 4-point Hikami boxes with a
few-impurity ladder in-between.
Several examples are highlighted by grey boxes in \Figs{fig:Hikami6}(b,c).
It is not \emph{a priori} clear whether such diagrams belong to the diagram
shown in \Fig{fig:GCE2}(b) or \Fig{fig:GCE2}(e). However, keeping
them only in the diagram \Fig{fig:GCE2}(b) allows us to maintain the
electroneutrality in all two-loop diagrams.
The total result for $H_6^{\rm (b)}$  is obtained by
summing 40 generated diagrams. The 6-point Hikami boxes of
\Figs{fig:GCE2}(c,d) can be calculated analogously.

Before presenting the final answer, we would like to discuss 
how to reinstate the finite dephasing rate in the equations. First, $ \, \gamma
\, $ must be included as a mass term in all Cooperon propagators.
Second, when calculating the Hikami box $H_4^{\rm (g2)}$ of \Fig{fig:GCE2}(g), only the number of coherent modes has to be conserved.
The latter is in contrast to all other Hikami boxes, which obey the usual electroneutrality condition,
i.e., the conservation of the {\it total} number of particles.
Hence, $H_4^{\rm (g2)}$ is the only Hikami box
of the two-loop calculations which is sensitive to dephasing of the
Cooperons.
This statement can be checked directly with the help of the model of magnetic impurities.
Introducing a slightly reduced scattering rate for all elastic collisions in the
particle-particle channel, $1/\tau \to 1/\tau - \gamma_{\rm mi}$, where $\gamma_{\rm mi} \ll 1/\tau$, and keeping $1/\tau$ for collisions in
the particle-hole channel, we observe that the Cooperon acquires the mass $\gamma_{\rm mi}$
since magnetic scattering breaks time-reversal symmetry. Hence, magnetic scattering rate is
similar to the dephasing rate; they both provide consistent infrared cut-offs for Cooperons.
Applying the algorithm described above, we find that, among all the two-loop diagrams in
\Fig{fig:GCE2}, the rate $\gamma_{\rm mi}$ appears only in the expressions for Cooperons and in
the Hikami box $H_4^{(g2)}$. In the latter case, it leads to changing $D \vQa^2$
to $D \vQa^2 + \gamma_{\rm mi}$. Using the analogy between magnetic scattering
and dephasing, we conclude that $\gamma$ enters $H_4^{(g2)}$ in the same way. 

Omitting lengthy and tedious algebra which will be published elsewhere,
together with a detailed proof of the validity of our method and an  analysis of the IR
cut-off in systems with magnetic impurities, the answer for $ \, \delta\chi_{\rm GCE}^{(2)} \, $ reads:
\begin{align}\label{eq:GCE2}
  & \delta\chi_{GCE}^{(2)}(\vq,\wo)
  = \frac{1}{(2\pi)^2 \rho_0 V^2}
    \frac{2 i\wo D\vq^2}{(D\vq^2-i\wo)^2} \sum_{\vQa,\vQb}
  \Big[ \\\nonumber
   &   P_c(\vQa,\wo) P_c(\vQb,\wo) \
       \left( \frac{D\vq^2+i\wo}{D\vq^2-i\wo} +\frac{4 D (\vq\vQa)(\vq\vQb)/\vq^2}{D(\vq+\vQa+\vQb)^2-i\wo} \right) \\\nonumber
   & + P_d(\vQa,\wo) P_d(\vQb,\wo) \
       \left( \frac{2D [\vq(\vq+\vQa+\vQb)]^2/\vq^2}{D(\vq+\vQa+\vQb)^2-i\wo} - 1 \right) \\\nonumber
   & + P_c(\vQa,\wo) P_d(\vQb,\wo) \
       \left( 2 + 2i\wo \, P_c(\vQa,\wo) \right)
   \ \Big] \, .
\end{align}

To conclude this section, we would like to note that our method of dressing the Hikami
boxes goes far beyond the initial ideas of \Ref{Hastings_Inequivalence_1994}.
It is a very powerful and generic working tool which can be extended to even more
complicated diagrams, including higher loop corrections, and nontrivial physical
problems. For example, our method can be straightforwardly used to describe mesoscopic systems in
the ballistic regime, cf. \Ref{Zala_Interaction_2001}. Therefore, the
diagrammatic approach presented above is complimentary to the diffusive nonlinear $ \, \sigma$-model which fails to yield ballistic results.
One can invent alternative digrammatic tricks which help to avoid the complexity
of the Hikami boxes with scalar vertices. For instance,  the density response function
can be obtained by calculating the current response function (averaged conductivity) first
and then using the continuity equation. In the latter approach, the dressed scalar vertices
are replaced by undressed vector ones, which greatly simplifies the calculation\cite{Gorkov_Conductivity_1979}. However,
this method cannot describe the full $\vq$-dependence of $ \, \chi $, which is crucial for
the polarizability. We have checked that both approaches give the same results in the
small-$\vq$ limit.

\subsection{Canonical ensemble}
\label{sect:DensityResponseFunction-CE}

In this section, we study the disorder average of the density response
function $\chi$ in the CE, where the number of particles $N$ is fixed in
each sample. Let us first discuss the properties of the statistical
ensemble which corresponds to the experimental measurements of the polarizability,
such as the experiment discussed in \Sect{sect:ComparisonExperiment}.
We are mainly interested in the behavior close to the 0D regime, where
 due to $ \, \tPhi \ge 1/ \ETh$, there is no self-averaging. Instead, the
disorder average is usually realized by an ensemble average.
The samples from the ensemble differ in impurity configuration {\it and}
can have slightly different particle number. At $T=0$ (in the ground state)
all single-particle levels below the Fermi level $\eF$ are occupied.
However, one cannot fix  $ \, \eF \, $ for the whole ensemble due to
randomness of the energy levels  {\it and} due to the fluctuations of
$ \, N \, $ from sample to sample. This can be taken into account by
introducing an $\eF$ which fluctuates around the typical value $ \, \mu^0 $;\cite{Lehle_Canonical_1995}
$ \, \mu^0 $ fixes the mean value
of $ \, N \, $ in the entire ensemble. It has been shown that such
ensembles of isolated disordered samples with fluctuating
$\eF$ can be described by the so-called Fermi-level pinning ensemble,
\cite{Lehle_Canonical_1995,Altland_Canonical_1992}
which is realized as follows:
(i) the Fermi-energy is pinned to an energy level $\ep_\vk$, such that
$ \eF = \ep_\vk + 0$.
(ii) the level $\ep_\vk$ is sampled from a weight function $P(\ep_\vk)$,
which is centered at
$ \eFtyp $ and is normalized: $ \, \int P(\ep) {\rm d} \ep = 1 $.
The support of $P(\ep_\vk)$ should be much smaller than $\eFtyp$ but
much larger than $\Delta$. The correlations resulting from fixing
$ \, N \, $ in the given sample are subsequently reduced to the
additional correlations induced by disorder with the help of the
following procedure: The expression for the density response function
averaged over the fluctuating Fermi energies and over disorder reads:
\begin{align}\label{eq:chiCanon}
  \overline{\chi}(\vq,\wo)
  = & \frac{1}{\sum_\vk \overline{P(\ep_\vk)}}
      \sum_\vk \overline{P(\ep_\vk) \chi_{\ep_\vk}(\vq,\wo)} \, .
\end{align}
In \Eq{eq:chiCanon} we have assumed that the numerator and denominator
can be averaged over disorder independently, see the discussion in
\Ref{Lehle_Canonical_1995}. Since the averaged density of states depends
only weakly on disorder\cite{Montambaux_Book_2007} and is almost
constant on the support of $ \, P $, the denominator of \Eq{eq:chiCanon}
can be simplified
\begin{align}\label{eq:chiCanonDenom}
  \sum_\vk \overline{P(\ep_\vk)} = V \int_{-\infty}^{+\infty} \dx{E} P(E) \overline{\rho_E}
  \approx \rho_0 V \, .
\end{align}
Inserting \Eq{eq:chiMu} and  \Eq{eq:chiCanonDenom} into \Eq{eq:chiCanon},
we find the disorder averaged density response function in the CE:
\begin{align}\label{eq:ChiC}
  & \overline{\chi}(\vq,\wo)
  =  \frac{1}{\rho_0}
      \int_{-\infty}^{+\infty} \dx{E} P(E)
      \overline{\rho_{E} \, \chi_{E}(\vq,\wo)} =
              \overline{\chi}_{\mu} + \delta \chi_{CE} \, .
\end{align}
The loop-expansion of $ \, \overline{\chi}_{\mu} \, $ was calculated in
the previous section. The quantity $ \, \delta \chi_{CE} \, $ describes
additional contributions resulting from fluctuations of $ \, \eF $.
It is governed by the irreducible part of the integrand:
\begin{eqnarray}\nonumber
  \delta \chi_{CE} & \equiv  & \frac{1}{\rho_0}
      \int_{-\infty}^{+\infty} \dx{E} P(E)
      \left(
        \overline{\rho_{E} \, \chi_{E}(\vq,\wo)} - \overline{\rho_E} \, \overline{\chi_{E}(\vq,\wo)}
      \right) \\\label{eq:ChiCE}
     & \simeq & \frac{1}{\rho_0}
      \left(
        \overline{\rho_{E} \, \chi_{E}(\vq,\wo)} - \rho_0 \, \overline{\chi_{E}(\vq,\wo)}
      \right) \, .
\end{eqnarray}
In \Eq{eq:ChiCE}, we have assumed that the disorder-averaged quantities
are (almost) independent of the absolute values of the particle energies.
As a result, the exact form of the weight function
$P(\ep_\vk)$ is not important.
Let us now derive the leading contribution to $ \, \delta \chi_{CE} $.

\begin{figure}[t]
  \begin{center}
	\includegraphics[width=0.99\columnwidth]{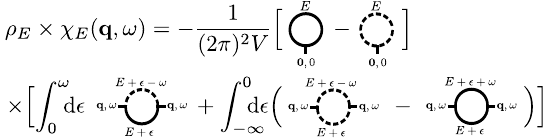}
  \end{center}
	\caption{Diagrammatic representation of the term
	$\rho_{E} \, \chi_{E}(\vq,\wo)$ from \Eq{eq:ChiCE} before impurity
	averaging, cf. \EqsTwo{eq:chiMu}{eq:defRho}.}
	\label{fig:CanonicalStructure}
\end{figure}

Diagrammatically, the additional factor $ \, \rho_{E} \, $ in \Eq{eq:ChiCE}
is represented as a closed fermionic loop with a vertex between two
(disorder averaged in further calculations) GFs which have the same
retardation, energy and momentum, see \Fig{fig:CanonicalStructure}.
Following \Ref{Smith_Spectral_1998}, we greatly reduce the number of
possible diagrams in \Eq{eq:ChiCE} by generating this vertex with the
help  of an additional energy derivative:
\begin{align}\label{eq:GFTrick}
  \overline{G}^{R/A}_{\ep}(\vk)^2
  = -\lim_{\lambda \to 0} \dfrac{\partial}{\partial \lambda}
                          \overline{G}^{R/A}_{\ep+\lambda}(\vk) \, .
\end{align}

\begin{figure}[t]
  \begin{center}
	\includegraphics[width=0.99\columnwidth]{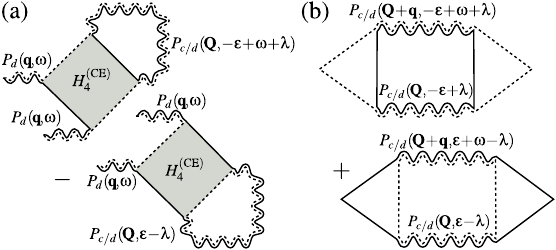}
  \end{center}
	\caption{One-loop diagrams which contribute to the disorder averaged
	         $ \, \delta \chi_{CE} $, \Eq{eq:ChiCE}, before taking the
	         derivative $\partial/\partial \lambda$, cf. \Eq{eq:GFTrick}.
	         Both 4-point Hikami boxes in (a) are given by
	         $H_4^{\rm (CE)} = 2 \pi \rho_0 \tau^4 ( D\vq^2 - i\wo )$.
          }
	\label{fig:CE1}
\end{figure}

After disorder averaging, we find two types of one-loop diagrams
which contribute to $ \, \delta \chi_{CE} $, see  \Fig{fig:CE1}: (i)
the diagrams in \Fig{fig:CE1}(a) are obtained by pairing the closed
loop with the $G^R G^A$ terms of $\chi_E$ (first term of the second
line of \Fig{fig:CanonicalStructure}); (ii) the diagrams of \Fig{fig:CE1}(b)
result form pairing with the $G^R G^R$/$G^A G^A$ terms (second and
third term). Furthermore, 4 more diagrams can be constructed where
Cooperon propagators are replaced by diffuson ones.

The double-counting problem does not appear in the diagrams in
\Fig{fig:CE1}(a), which contain  4-point Hikami boxes. Therefore, the
method which we used for the GCE diagrams is not needed here.
The only subtle issue in their calculation is that the diagrams are
small if the closed loop, $  \, \rho_E $, is connected to the bubble, $ \, \chi_E $,
by only one single impurity line.  Thus, at least two such connections
must be taken into account either in the ladder (which starts then from
two impurities) or in the ladder (which can start from one impurity)
{\it and} the particular dressing of the Hikami box which connects $ \rho_E $ to $ \chi_E $.
Furthermore, the 4-point Hikami box in \Fig{fig:CE1}(a) does not acquire a
dependence on dephasing rate $\gamma$, which can be checked with the help of the model of magnetic impurities
discussed before \Eq{eq:GCE2}. As a result, $\gamma$ has to be included only as
a mass term in the connected Cooperon.

Summing up all parts and calculating the auxiliary derivative, \Eq{eq:GFTrick},
we obtain the one-loop answer for $ \, \delta\chi_{CE} $:
\begin{align}\label{eq:CE1}
  & \delta\chi_{CE}^{(1)}(\vq,\wo) =
  \frac{2}{(2 \pi)^2 \rho_0 V^2} \sum_{\alpha=c,d} \sum_\vQ \\\nonumber
  & \quad \times \Big[
      \tfrac{i\wo}{D\vq^2 - i\wo} P_\alpha(\vQ,\wo) P_\alpha(\vQ,0)
      + P_\alpha(\vQ+\vq,\wo) P_\alpha(\vQ,0)
   \Big] \, .
\end{align}
Electroneutrality is restored in \Eq{eq:CE1} after summing all
the diagrams of \Fig{fig:CE1}. Thus, all contributions, \EqsThree{eq:GCE1}{eq:GCE2}{eq:CE1},
obey the electroneutrality condition; therefore, $ \, \overline{\chi}(\vq=0,\wo) = 0 $.

Note that the one-loop contribution $ \, \delta\chi_{CE}^{(1)} $,
\EqPlain{eq:CE1} is of the same order in $(\Delta/\max(\wo,\gamma))$,
$(\Delta/\wo)$ or $(\Delta/D\vq^2)$ as the two-loop contribution
$ \, \delta\chi_{GCE}^{(2)} $, \Eq{eq:GCE2}.
As a result, the differences between GCE and CE disappear at large frequencies $\wo \gg \Delta$,
in agreement with \Ref{Noat_Polarizability_1996}.
At smaller frequencies and weak dephasing, $\max(\wo,\gamma) \lesssim \Delta$, 
$ \, \delta\chi_{GCE}^{(2)} $
is needed to analyze the difference between the GCE and the CE for energies of
the order of $ \, O(\Delta) $.
In the following, we will often refer to
$\delta\chi_{GCE}^{(1)} $ as the result from ``1st order'' perturbation
theory, and $\delta\chi_{GCE}^{(1)}+\delta\chi_{GCE}^{(2)}+\delta\chi_{CE}^{(1)}$
(or $\delta\chi_{GCE}^{(1)}+\delta\chi_{GCE}^{(2)}$) as the result from
``2nd order' perturbation theory for isolated (or connected) systems.

\section{Quantum corrections to the polarizability}
\label{sect:QuantumCorrections}

The quantum corrections to $ \, \alpha \, $ can be found after inserting
the decomposition $ \, \chi = \chi_0 + \delta \chi \, $ into \Eq{eq:AlphaInitital} and
expanding the density response function in the RPA, $ \, \chi / \ep $,
in  $ \, \delta \chi $. Note that the latter can contain $ \, \delta \chi_{GCE}^{(1,2)} \, $
and $ \, \delta \chi_{CE}^{(1)} \, $ depending on the ensemble which we
consider and on the accuracy of the loop-expansion. This expansion up to
terms of order $ \, O(\delta \chi)^2 \, $ yields:
\begin{align}\label{eq:ChiRPAExp}
  \frac{\chi(\vq,\wo)}{\ep(\vq,\wo)}
  \approx \frac{\chi_0(\vq,\wo)}{\ep_0(\vq,\wo)}
          \Bigg[ 1
  & + \frac{1}{\ep_0(\vq,\wo)} \frac{\delta \chi(\vq,\wo)}{\chi_0(\vq,\wo)} \\\nonumber
  & + \frac{1-\ep_0(\vq,\wo)}{\ep_0(\vq,\wo)^2} \left(\frac{\delta \chi(\vq,\wo)}{\chi_0(\vq,\wo)}\right)^2
          \Bigg] \, ,
\end{align}
where $\ep_0(\vq,\wo) = 1 - 2U(\vq) \chi_0(\vq,\wo)$.
To separate the frequency dependence due to classical diffusive screening from
the frequency dependence of the quantum corrections, it is convenient to rewrite
\Eq{eq:ChiRPAExp}  as follows:
\begin{align}\label{eq:ChiRPASF}
  \frac{\chi(\vq,\wo)}{\ep(\vq,\wo)}
  \approx \
   \rho_0 S(\vq,\wo) \bigg[&  1
  + 2 \frac{S(\vq,\wo)}{g(|\vq|^{-1})} F(\vq,\wo) \\\nonumber
  & + 8 \, U(\vq) \chi_0(\vq,\wo) \frac{S(\vq,\wo)^2}{g(|\vq|^{-1})^2} F(\vq,\wo)^2
  \bigg] \, .
\end{align}
Here we have introduced two dimensionless functions:
\begin{align}\label{eq:defS}
  S(\vq,\wo)
  \equiv \left(\!\! 1 \!-\! 2 U(\vq)\rho_0 \!-\! \frac{i\wo}{D\vq^2} \!\!\right)^{\!\!\!-1} \, ,
\end{align}
which describes classical diffusive screening, and
\begin{align}\label{eq:defF}
  F(\vq,\wo)
  \equiv & \frac{(D\vq^2-i\wo)^2}{D\vq^2} \pi V \, \delta \chi(\vq,\wo) \, ,
\end{align}
which describes the quantum corrections to $ \, \chi $.
$ \, g(L) \, $ denotes the dimensionless conductance of a diffusive system
of size $ L $:
\begin{align}\label{eq:defG}
   g(L) \equiv 2\pi \, \ETh\!(L) / \Delta \, , \quad \ETh\!(L) = D/L^2 \, .
\end{align}
\EqsFromTo{eq:ChiRPASF}{eq:defF} together with \EqsThree{eq:GCE1}{eq:GCE2}{eq:CE1}
are the first major results of this paper.
The quantum corrections $\Delta \alpha$ are obtained by substituting the
terms $\sim F$ and $\sim F^2$ of \Eq{eq:ChiRPASF} into \Eq{eq:AlphaInitital}
and summing over $\vq$.
We remind the reader that the zero mode does not contribute to the polarizability
due to electroneutrality $ \, \chi(\vN,\wo)=0 $ and, therefore, we can assume
$ \, |\vq| \ne 0 \, $ in \Eq{eq:ChiRPASF}. The typical momenta which govern
the sum in \Eq{eq:AlphaInitital} are $ \, |\vq| \sim 1/L \, $ since the external
potential $\phi_{\rm ext}$ varies on the scale of the sample size $L$. But
we will keep $ \, \vq \, $ below for generality.

\section{Comparison to RMT+$\sigma$-model}
\label{sect:ComparisonRMTSigma}

Let us now compare the results of our perturbative calculations with those of
\Ref{Blanter_Polarizability_1998}
which are obtained from a combination of the RMT approach and the nonlinear
$\sigma$-model. The latter will be referred to as ``RMT+$\sigma$-model''.
This comparison requires an assumption $ \, \ETh(L) \gg \max(\Delta,
\wo,\gamma) \, $ which in particular means $ \, g(L) \to \infty $. In this
limit, the term $ \, \sim F^2 \, $ in \Eq{eq:ChiRPASF} acquires an additional
smallness (which can be estimated as $ \, O(1/g(L)) $) and can be neglected
while the term $ \, \sim F^1 \, $ becomes independent of $\vq$.
Next, we keep only the zero mode contributions in all sums over internal
momenta in the expressions for $ \, \chi_{GCE}^{(1,2)} \, $ and $ \, \delta
\chi_{CE}^{(1)} \, $ and consider the difference of $ \, F \, $ calculated for
unitary and orthogonal ensembles: $ \, \delta_B F(\wo) = F(\wo,B \to \infty) -
F(\wo,0) $, where $B$ is the strength of an external magnetic field.
The terms which contain only diffusons are canceled in $ \,
\delta_B F $.

Using \EqsThree{eq:GCE1}{eq:GCE2}{eq:CE1}, we obtain
\begin{align}\label{eq:DeltaFSimple}
   \delta_{B} F(\wo, & g \to \infty) =  \\\nonumber
  &
    \underbrace{-\frac{i\wo}{\gamma-i\wo}}_{\delta \chi_{GCE}^{(1)}} - \frac{\Delta}{2\pi}
    \Big[ \underbrace{
            \frac{i\wo-2\gamma}{(\gamma-i\wo)^2}
          }_{\delta \chi_{GCE}^{(2)}} +
          \underbrace{
               \frac{2 \gamma}{\gamma(\gamma - i\wo)}
          }_{\delta \chi_{CE}^{(1)}}
    \Big] \, .
\end{align}
Subscripts under the braces explain the origin of the corresponding terms.
The last term must be taken into account only in the CE. The
counterpart of \Eq{eq:DeltaFSimple} obtained from RMT+$\sigma$-model in
\Ref{Blanter_Polarizability_1998} reads:
\begin{align}\label{eq:DeltaFRMTSigma}
  & {\rm RMT+\sigma:} \quad
    \delta_{B} F(\wo)
    = 1 + \int_{+0}^{\infty} \frac{\dx{\ep}}{\Delta}
    \left( \frac{1}{\ep-\wo} + \frac{1}{\ep+\wo} \right) \\\nonumber
  & \times \left[
     \overbrace{
      \underbrace{
        \ep  \, \delta_B R_2(\ep)
      }_{GCE} +
        \Delta \, \delta_B R_2(\ep) +
        \int_{+0}^{\ep-0} \dx{\ep_1} \, \delta_B \tilde{R}_3(\ep,\ep_1)
      }^{CE} \right] \, .
\end{align}
Here $R_{2,3}$ are the usual (dimensionless) two- and three-level spectral
correlation functions, $\tilde{R}_3(\ep,\ep_1) = R_3(\ep,\ep_1)- R_2(\ep) \, $,
and $ \,  \delta_B R_{2,3} $ denotes the difference of the correlation functions
without and with time-reversal symmetry.
We have marked in \Eq{eq:DeltaFRMTSigma} the relevance of different
terms for the GCE and the CE.

\begin{figure}[t]
  \begin{center}
	\includegraphics[width=0.99\columnwidth]{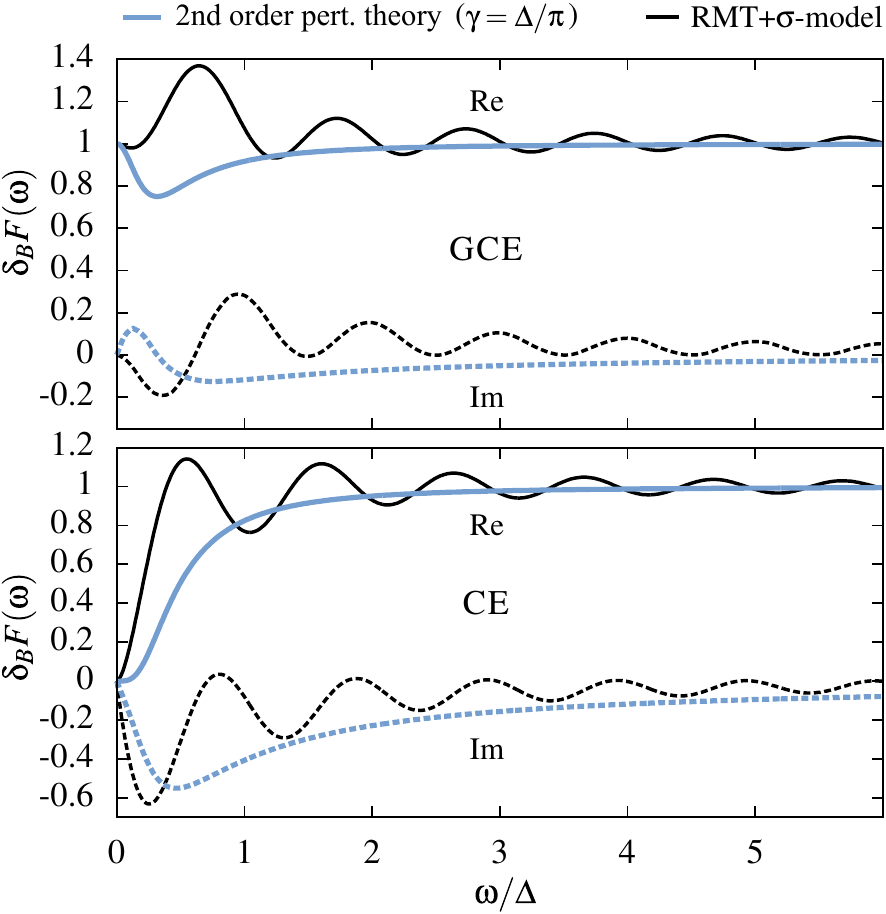}
  \end{center}
	\caption{(color online)
           The quantum corrections to the polarizability
	         in the limit $ \, \ETh\!(L) \gg \max(\Delta,\wo,\gamma) \, $
           for the GCE (upper panel) and the CE (lower panel).
           We compare real (solid lines) and imaginary (dashed lines) parts
	         the function $\delta_B F$ obtained from 2nd order perturbation
	         theory, \Eq{eq:DeltaFSimple}, and from the RMT+$\sigma$-model,
	         \Eq{eq:DeltaFRMTSigma}.
	        }
	\label{fig:RMT}
\end{figure}

We remind the reader that the RMT+$\sigma$-model results are valid for
$ \, \gamma = 0 \, $ and cannot straightforwardly describe a
$\gamma$-dependence, while our perturbative  result, \Eq{eq:DeltaFSimple},
is valid only if $  \Delta  \lesssim {\rm max} \left( \gamma, \wo \right) $.
To resolve this issue, one should set in \Eq{eq:DeltaFSimple} $\gamma \sim \Delta$.
\Eq{eq:DeltaFSimple} yields $ \, \delta_B F(\wo \to 0, g \to \infty) = \Delta/(\pi \gamma) $
for the GCE. Therefore, we have chosen $ \, \gamma=\Delta/\pi \, $ to
ensure the correct limit $ \, \delta_B F(\wo \to 0, g \to \infty) \big|_{GCE} = 1 $.

The comparison of the results obtained from RMT+$\sigma$-model and from
the perturbative calculations  are shown in \Fig{fig:RMT} for the GCE and
the CE. Apart from the oscillations in the RMT+$\sigma$-curves, whose origin is nonperturbative, the
agreement is excellent. The asymptotic limits are fully recovered in the
perturbative calculations:
(i)~$ \delta_B F(\wo \gg \Delta,g\to\infty) \to 1 \, $ for the both ensembles;  
(ii)~$ \delta_B F(\wo~\to~0,g\to\infty) \to 0 \, $ in the CE due to cancellation of
$ \, \delta \chi^{(2)}_{GCE} \, $ and $ \, \delta \chi^{(1)}_{CE} $.
The latter property of the CE holds true at
any $\gamma$ in 1st \emph{and} 2nd order perturbation theory.
In the GCE, on the other hand,
the quantum corrections remain finite for $\wo \to 0$ in 2nd order perturbation theory,
in full agreement with the nonperturbative results of \Ref{Blanter_Polarizability_1998}
\cite{Noat_Comment}.

We conclude this section by noting that the perturbation theory
is able to reproduce the results of the RMT+$\sigma$-model with good qualitative agreement,
which is the second major result of our work.

\section{Polarizability of an ensemble of rings}
\label{sect:ComparisonExperiment}

The experiments described in \RefsTwo{Deblock_Experiment_2000}{Deblock_Experiment_2002}
were done on a large number of disordered metallic rings. The rings were
etched on a 2D substrate and were placed on the capacitative part of a
superconducting resonator, where a spatially homogeneous in-plane electric
field $\vE(\wo)$ acted on them. In terms of the coordinate along the ring,
$x\in[0,2 \pi R]$, where $R$ is the ring radius, the external electric
potential of this field is
$\phi_{\rm ext}(x,\wo) = |\vE(\wo)| R \cos(x/R) + \phi_{\rm ext}^{(0)}$,
and its Fourier transform reads
\begin{equation}\label{eq:PhiExtRing}
  \phi_{\rm ext}(q,\wo)
     = -|\vE(\wo)| R^2 \pi \left[ \delta_{q,1/R} + \delta_{q,-1/R} \right]
     +  \phi_{\rm ext}^{(0)} \cdot \delta_{q,0} \, .
\end{equation}
The constant shift of the potential $ \, \phi_{\rm ext}^{(0)} \, $ does not contribute
to the polarizability. Therefore, the sum in \Eq{eq:AlphaInitital} involves only
two modes, $q=1/R$ and $q=-1/R$, which yield
\begin{eqnarray}\nonumber
  \alpha(\wo) & = & \frac{4 e^2}{|\vE(\wo)|^2}
  \frac{1}{2 \pi R} \phi^2_{\rm ext}(q,\wo)
  \frac{\chi(q,\wo)}{\ep(q,\wo)} \Big|_{q=1/R} \\\label{eq:AlphaRing}
   & = & 2 \pi e^2 R^3 \frac{\chi(1/R,\wo)}{\ep(1/R,\wo)} \, .
\end{eqnarray}
In \Eq{eq:AlphaRing}, we have taken into account the symmetry of the summand
under the inversion $ \, q \to - q $.

The Coulomb potential in quasi-1D is given by
\begin{equation}\label{eq:Q1d-U}
   U(q) = 2e^2 \ln(|qW|) \, , \quad |qW| \ll 1 \, ;
\end{equation}
where $ \, W \ll R \, $ is the width of the ring. Inserting \Eq{eq:Q1d-U}
into \Eq{eq:defS}, we find the screening function of the quasi-1D ring
at $ \, q=1/R $:
\begin{align}\label{eq:SRing}
  S(1/R,\wo)
  & = \left( 1 + (\kappa W) \ln(R/W)/\pi
      - \frac{i\wo}{\ETh\!(R)} \right)^{-1} \\\label{eq:SRingApprox}
  & \overset{\kappa W \gg 1}{\approx} \frac{\pi}{(\kappa W) \ln(R/W)} \equiv S_0 \ll 1 \, .
\end{align}
We have introduced the 2D Thomas-Fermi screening vector,
$\kappa = 4 \pi e^2 \rho_0/W \, $ with $ \, \rho_0 \, $ being the
\emph{quasi-1D} density of states, see e.g. \Ref{Montambaux_Book_2007},
and assumed sufficiently strong screening, $\kappa W \gg 1$, such that
$ \, S \, $ reduces to the $ \, \wo $-independent constant $ \, S_0 $.
This agrees with the experiment where one can estimate $(\kappa W) \ln(R/W) \approx 18$.
Therefore, we focus below only on the limit of strong screening.
Note that in this limit, the product $ \,  U(1/R,\wo) S(1/R,\wo) \, $
can be also simplified
\begin{align}\label{eq:StrongScreening}
  U(1/R,\wo) \, S(1/R,\wo) \approx - 1/2\rho_0 \, .
\end{align}

The classical part of the polarizability comes from inserting
the leading term of the expansion \EqPlain{eq:ChiRPASF} into \Eq{eq:AlphaRing}:
\begin{align}\label{eq:AlphaRingClass}
  \alpha_0 \simeq  2 \pi e^2 R^3 \, \rho_0 S_0 = \frac{\pi R^3}{2 \ln(R/W)} \, .
\end{align}
Using \EqsTwo{eq:chi0}{eq:StrongScreening} in \Eq{eq:ChiRPASF}, and
inserting the result into \Eq{eq:AlphaRing}, we obtain the quantum
corrections to the polarizability up  to the term $ \, \sim (F/g)^2 $:
\begin{align}\label{eq:DeltaAlphaRing}
  \frac{\Delta \alpha (\wo)}{2 S_0 \alpha_0}
  \approx \frac{F(R^{-1},\wo)}{g(R)}
   - 2  \frac{\ETh\!(R)}{ \ETh\!(R) - i\wo}
   \left( \frac{F(R^{-1},\wo)}{g(R)} \right)^{\!\!2} .
\end{align}
Let us regroup the terms in \Eq{eq:DeltaAlphaRing} to single out the
terms of 1st and 2nd order perturbation theory:
\begin{align}\label{eq:DeltaAlphaRingF}
  \frac{\Delta \alpha (\wo)}{2 S_0 \alpha_0}
  \approx & \frac{1}{g(R)} \left( F^{(1)}(1/R,\wo) + F^{(2)}(1/R,\wo) \right)
\end{align}
with
\begin{align}\label{eq:FFinal1}
  F^{(1)}(1/R,\wo) = (2\pi^2 R) \frac{(\ETh\!(R)-i\wo)^2}{\ETh\!(R)} \delta \chi_{GCE}^{(1)}(1/R,\wo) ,
\end{align}
and
\begin{align}\label{eq:FFinal2}
  & F^{(2)}(1/R,\wo)  = \\\nonumber
  & \ \ (2\pi^2 R) \frac{(\ETh\!(R)-i\wo)^2}{\ETh\!(R)}
    \left( \delta \chi_{GCE}^{(2)}(1/R,\wo) + \delta \chi_{CE}^{(1)}(1/R,\wo) \right) \\\nonumber
  & \ \ + \frac{2}{g(R)} (2 \pi^2 R)^2 \frac{(\ETh\!(R)-i\wo)^3}{\ETh\!(R)}
    \left(  \delta \chi_{GCE}^{(1)}(1/R,\wo) \right)^2 \, .
\end{align}

We emphasize that all three parts of the density response function,
$ \, \delta \chi_{GCE}^{(1,2)} \mbox{ and } \delta \chi_{CE}^{(1)} $,
are generically important for the theoretical description of the
experimental data with the help of \Eq{eq:DeltaAlphaRingF} if the rings
are isolated. Having obtained \EqsThree{eq:GCE1}{eq:GCE2}{eq:CE1} (and
\Eq{eq:DeltaFSimple} for the limit $g\to\infty$) and
\EqsFromTo{eq:DeltaAlphaRing}{eq:FFinal2}, we are now in the
position to analyze different  options to fit the experimental data.
\RefsTwo{Deblock_Experiment_2000}{Deblock_Experiment_2002} focused on the
$T$-dependence of the real part of the quantum corrections, thus,
in the following we will concentrate on ${\rm Re} \Delta \alpha$.

\begin{figure}[t]
  \begin{center}
	\includegraphics[width=0.99\columnwidth]{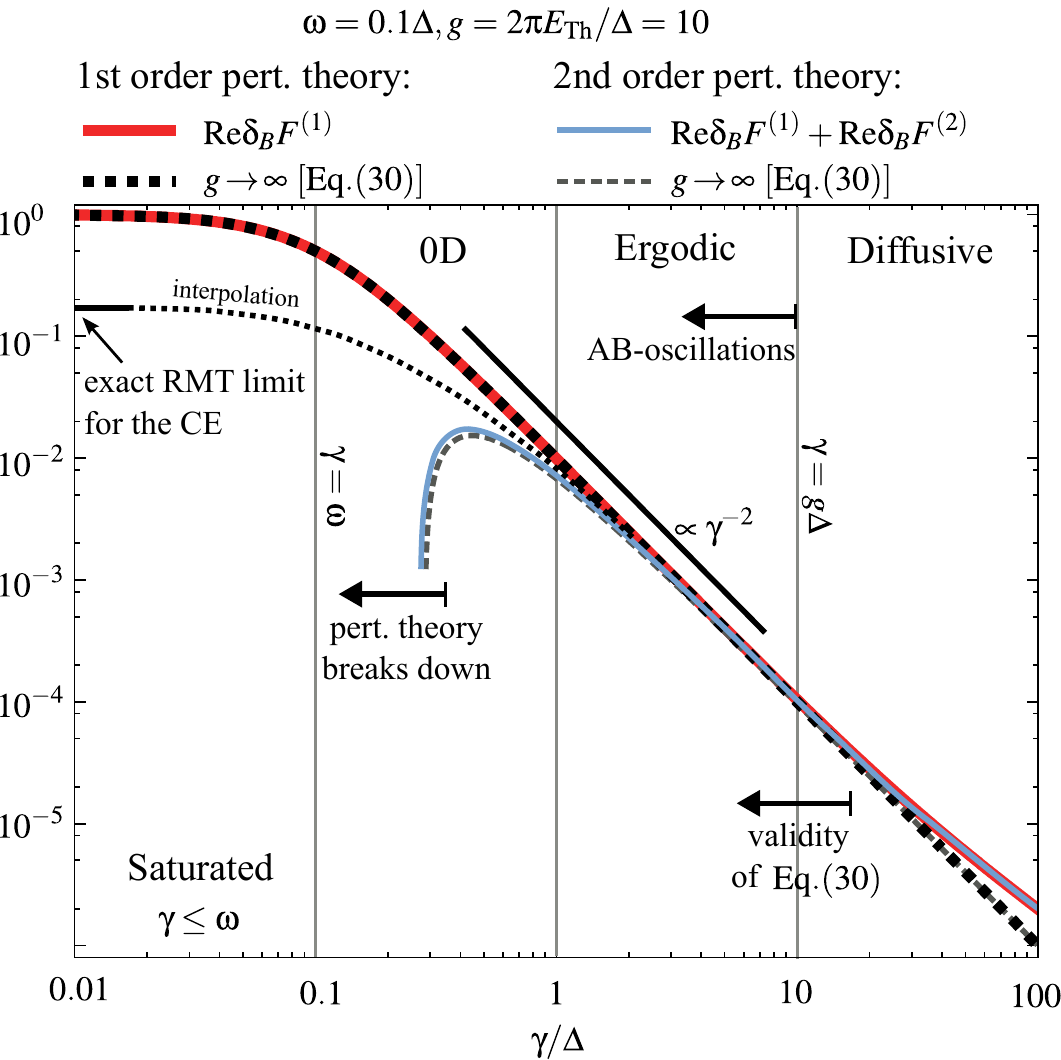}
  \end{center}
	\caption{(color online) Comparison of perturbative 1st order, $\delta_B F^{(1)}$, 2nd order,
 	         $\delta_B F^{(1)} + \delta_B F^{(2)}$, and interpolated (to the RMT+$\sigma$-model limit) results for the
	         quantum corrections to the polarizability in the parameter range $\wo < \Delta < \ETh $.
	        }
	\label{fig:Result}
\end{figure}

\begin{figure}[t]
  \begin{center}
  \includegraphics[width=0.99\columnwidth]{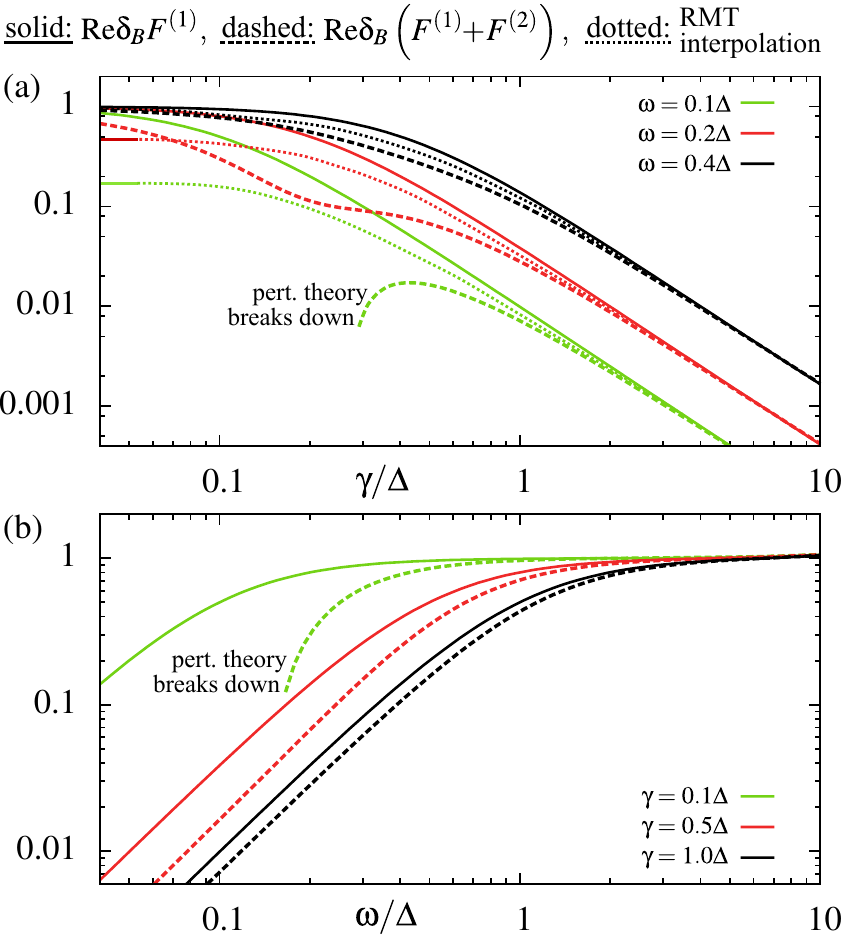}
  \end{center}
	\caption{(color online) Comparison of perturbative 1st order, $\delta_B F^{(1)}$, 2nd order,
 	         $\delta_B F^{(1)} + \delta_B F^{(2)}$, and interpolated (to the RMT+$\sigma$-model limit) results for the
	         quantum corrections to the polarizability (a) as a function of $\gamma$ for different values of $\wo$ and
	         (b) as a function of $\wo$ for different values of $\gamma$.
	        }
	\label{fig:Validity}
\end{figure}

The crossover to 0D dephasing occurs when $\gamma$ decreases below $\Delta$.
We expect that the ideal parameter range to study this crossover
experimentally in the CE is $ \, \wo < \Delta < \ETh $. However, it is
important that the conductance should be only moderately large, since
$\Delta \alpha$ is suppressed in the case of extremely large $ \, g $,
cf. \Eq{eq:DeltaAlphaRing}; and the frequency should not be too small, since
the quantum corrections to the polarizability of isolated systems are
suppressed in the static limit, see \Fig{fig:RMT}. Let us first discuss
our general expectations for this parameter range, which are illustrated
in \Fig{fig:Result}. The simplest regime is
$ \, 1 \lesssim \gamma/\Delta \lesssim g \, $ where the loop-expansion
can be justified and the difference between the GCE and the CE is negligible.
Keeping only the  leading term, we obtain a power law for the dependence
of $ \, \Delta \alpha \, $ on $ \, \gamma $. This power law can be derived
straightforwardly after noting that, in the range $(\gamma,\wo)/\Delta \ll g $,
one can use the approximation \Eq{eq:DeltaFSimple} and find
${\rm Re}\Delta \alpha \sim {\rm Re}\delta_B F(\wo) \sim \wo^2/\gamma^2$
for $\wo \ll \gamma$.

The subleading terms, which in particular describe the difference of the
GCE and the CE, are able to improve the theoretical answer for
$ \, \gamma \, $ being slightly smaller than $ \, \Delta $. However,
$ \, {\rm Re} \delta_B F^{(2)} \, $ (and, correspondingly, the difference
between the ensembles) is small at any $ \, \gamma \, $ for moderately small
frequencies, see the example $ \, \wo  = 0.4 \Delta \, $ in \Fig{fig:Validity}.
Therefore, $ \, \delta_B F^{(1)} \, $ suffices to fit the experiment at
$ \wo \gtrsim 0.4 \Delta$. The $ \, T $-dependence of
$ \, \Delta \alpha \, $ saturates to the value predicted by the
RMT$+\sigma$-model at $ \, \gamma \lesssim \wo $ which makes the range
of pronounced 0D dephasing ($\wo\lesssim\gamma\lesssim\Delta$) too
narrow even at $ \, \wo \simeq 0.4 \Delta $, thus, smaller frequencies
are needed. Of course, the perturbation theory is no longer valid if both
$ \, \wo \, $ and $ \, \gamma \, $ are small. In particular, when
$ \, F^{(2)} \, $ becomes of order of $ \, F^{(1)} \, $ it can lead to
changing the overall sign of $ \, {\rm Re}\delta_B(F^{(1)}+F^{(2)}) $,
see the cut of the lines in \Fig{fig:Result} marked ``pert. theory breaks down''. We believe that this sign
change is unphysical and, moreover, it contradicts the prediction of the
RMT+$\sigma$-model. Nevertheless, our calculations show that the power law,
which is obtained in the perturbative region from the leading correction,
can be extended well into the nonperturbative region
$ \wo \lesssim \gamma \lesssim \Delta$. This provides us with the unique
possibility to detect the crossover to 0D dephasing directly from the
amplitude of $\Delta \alpha$. It is in sharp contrast to the
quantum corrections to the conductivity, which always saturate at
$\gamma \lesssim \Delta$. \cite{Treiber_DimensionalCrossover_2009,Treiber_QuantumDot_2012}

Let us illustrate our unexpected statement with the help of \Fig{fig:Result}:
We know the exact value of $ \, \Delta \alpha \, $ in the limit
$\gamma \to 0$ from the RMT+$\sigma$-model and the correct behavior of
$ \, \Delta \alpha $ for $ \, \gamma \, $ being of order of (and slightly
below) $ \, \Delta $. Using these reference points, one can interpolate
the dependence $ \, \delta \alpha(\gamma) \, $ for the whole region
$ \, 0 < \gamma \lesssim \Delta $. Since the slope of the interpolated
curve is only slightly different from the perturbative one for
$ \, \, \gamma \ge 0.3 \Delta $, {\it the leading answer of perturbation
theory can be used to detect the crossover to 0D dephasing}. If the range
$ \, \gamma \ge 0.3 \Delta $ is not sufficient for unambiguously fitting the
experiment, the whole interpolated curve can be used instead.

The authors of \RefsTwo{Deblock_Experiment_2000}{Deblock_Experiment_2002}
used a superconducting resonator with fixed frequency
$ \, \omega \simeq 0.2 \Delta \simeq 17 {\rm mK} \, $ to measure
$ \, \Delta \alpha(T) \, $ of the rings. In the following we will apply
our theory to explain the experimental results of these papers.
We note that the qualitative difference in the slope of the curves obtained from the three options for fitting -- (i)
the interpolated curve, (ii) the result of 2nd order perturbation theory,
and (iii) the leading perturbative result -- becomes rather insignificant
at $ \, \omega \simeq 0.2 \Delta $ and $ \gamma \gtrsim 0.3 \Delta $, see \Fig{fig:Validity}(a).
The main difference between (i) and (iii) is that the saturation originates
at slightly larger $ \, \gamma \, $ than the leading perturbative result
would suggest.
Thus we can safely keep $ \, F^{(1)} \, $ and neglect
$ \, F^{(2)} \, $ to fit the data, which makes our task simpler
\cite{phi_0_oscillations}.

\begin{figure}[t]
  \begin{center}
	\includegraphics[width=0.99\columnwidth]{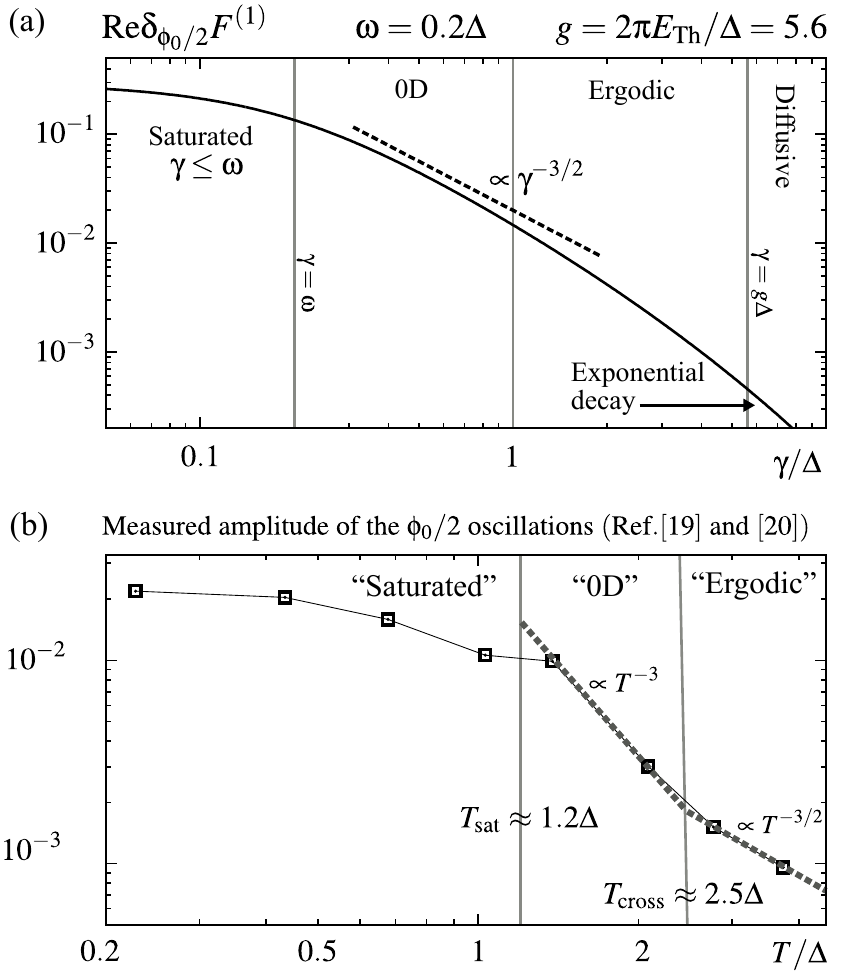}
  \end{center}
	\caption{Amplitude of the $\phi_0/2$ oscillations.
	         (a) Expected dependence on $\gamma$ from our theory, \Eq{eq:DeltaFGCEFlux}, for the
	         parameter range $\wo \ll \Delta \ll \ETh $. (b) Experimentally measured
	         data as a function of temperature and possible interpretation. Note that
	         the theory (see e.g. \RefsTwo{Treiber_DimensionalCrossover_2009}{Treiber_ThermalNoise_2011})
	         predicts $\gammaZD \propto T^{2}$ in the 0D regime, and $\gammaErg \propto T$
	         in the ergodic regime, therefore, the $\gamma^{-3/2}$ behavior indicated in (a)
	         encompasses both the $T^{-3}$ and $T^{-3/2}$ behavior seen in (b).
	        }
	\label{fig:TempFit}
\end{figure}

The experimental results for the ring polarizability can be distorted
because of a parasitic contribution  from  the resonator. The latter
has been filtered out in the experiment with the help of an additional
weak magnetic field $ \, B \, $ applied perpendicular to the rings, such
that $\Delta \alpha$ becomes a periodic function of the magnetic flux
through the ring. Measuring the $T$-dependence of the $\phi_0/2$
oscillations, cf. Fig.~9 of \Ref{Deblock_Experiment_2002}, allows one to
focus purely on the response of the rings.
Using \Eq{eq:GCE1} in \Eq{eq:defF}, we find
\begin{align}\label{eq:DeltaFGCE1}
  \Delta \alpha & \propto F^{(1)}(1/R,\wo) = i\wo \sum_\vQ P_c(\vQ,\wo) \\\nonumber
  & = i\wo L \int_0^{\infty} \dx{t} e^{i \wo t}
      \sum_n \frac{1}{\sqrt{4 \pi D t}}
      e^{-(nL)^2/4Dt} e^{i \theta n} e^{-\gamma \, t} \, ,
\end{align}
where $\theta = 4\pi \phi/\phi_0$ and $\phi$ is the flux through one ring,
and $ \, L=2\pi R$. Taking the Fourier transform and selecting the $\phi_0/2$
signal gives:
\begin{align}\label{eq:DeltaFGCEFlux}
  \delta_{\phi_0/2} F^{(1)}(1/R,\wo)
  = & \frac{i\wo \ {\rm exp}\left(-\sqrt{(\gamma - i\wo)/\ETh\!(L)} \right)}
           {\sqrt{\ETh\!(L)(\gamma - i \wo)}} \, .
\end{align}
The function $\delta_{\phi_0/2} F^{(1)}$ is shown in \Fig{fig:TempFit}(a).
It is similar to $\delta_B F^{(1)}$, cf. \Fig{fig:Result}, however, the
dependence of $\delta_{\phi_0/2} F^{(1)}$ on $\gamma$ is governed by a
$\propto \gamma^{-3/2}$ power law in the regime $\wo \ll \gamma \ll g\Delta$,
and in the regime $ g \Delta \ll \gamma$, the $\phi_0/2$ oscillations are
exponentially suppressed. The theory predicts a 0D dephasing rate,
$\gammaZD = a \Delta T^2/\ETh^2$,\cite{Sivan_QuasiParticleLifetime_1994} at low temperatures and an ergodic
dephasing rate, $\gammaErg = b \Delta T/\ETh$,\cite{Ludwig_AharonovBohm_2004,Texier_DiffusiveRing_2005} at higher temperatures,
where $a$ and $b$ are system-specific, dimensionless coefficients of
order $\sim 1$, see \RefsTwo{Treiber_DimensionalCrossover_2009}{Treiber_ThermalNoise_2011}.
The crossover between the two regimes occurs at a temperature $\Tcross= \tfrac{b}{a} \ETh $.
We expect that the saturation at $\gamma=\wo$ occurs in the 0D regime,
corresponding to a temperature $T_{sat}=\tfrac{1}{\sqrt{a}} \ETh \sqrt{\wo/\Delta}$.
Note that the conductance of each ring was rather small, $ \, g(L) \approx 5.6 \, $,
such that the Thouless energy $\ETh\!(L) \approx 0.9 \Delta $. Thus, depending
on the coefficients $a$ and $b$, $\Tcross$ and $\Tsat$ can be relatively
close to each other.

The experimental result for the $T$ dependence of the $\phi_0/2$ oscillations
is shown in \Fig{fig:TempFit}(b). The measurements were done
in the temperature interval $ \wo \simeq 0.2 \Delta \le T \le 4 \Delta $.
Based on the preceding discussion, we offer the following interpretation of
the data: At low temperatures $T\lesssim 1.2 \Delta$,
the quantum corrections depend only weakly on $T$ and are almost saturated.
At intermediate temperatures $1.2\Delta \lesssim T \lesssim 2.5 \Delta$ the slope
of the data is steep and consistent with 0D dephasing $\Delta \alpha(T) \propto \gammaZD^{-3/2} \propto T^{-3}$.
At higher temperatures $T \gtrsim 2.5 \Delta$, the slope of $\Delta \alpha(T)$ decreases and is
consistent with ergodic dephasing $\Delta \alpha \propto \gammaErg^{-3/2} \propto T^{-3/2}$.
The crossover temperatures, $\Tsat \simeq 1.2\Delta$ and $\Tcross \simeq 2.5\Delta$,
correspond to coefficients $a\simeq 0.1$ and $b \simeq 0.3$, which are close to
the values predicted in \Ref{Treiber_DimensionalCrossover_2009} ($a\simeq 0.04$ and $b\simeq 1$).
However, we stress that this interpretation is based only on very few data
points, and we do not claim that the experiment clearly shows a crossover
to 0D dephasing. Further experiments are needed to support this statement,
see \Sect{sect:conclusions}.

\section{Conclusions}
\label{sect:conclusions}

Understanding interference phenomena and dephasing in mesoscopic systems at
very low temperatures is a subtle issue which has provoked controversies
between different theoretical approaches \cite{Marquardt_Decoherence_2007},
as well as between theory and experiments \cite{Huibers_Dephasing_1998}.
Quantum transport experiments cannot give a certain answer to all questions
because of unavoidable distortions due to the coupling to the environment.
The response of isolated disordered samples, on the other hand, provides
a ``cleaner'' setup to study dephasing, and gives one the possibility to
settle long-lasting open questions.

We have studied the quantum corrections to the polarizability of isolated
disordered metallic samples aiming to improve the explanation of previous
experiments (\RefsTwo{Deblock_Experiment_2000}{Deblock_Experiment_2002}),
and to suggest new measurements, where the elusive 0D regime of dephasing
can be ultimately detected. Using the standard strategy of mesoscopic
perturbation theory, i.e. the loop-expansion in diffusons and Cooperons,
we have developed a theory, which (i) accounts for the difference between
connected (GCE) and isolated (CE) systems, and (ii) is able to describe
the low frequency response of disordered metals, taking into consideration
weak dephasing induced by electron interactions. We have shown that the
difference between the GCE and the CE appears only in the subleading terms,
therefore, we have extended the calculations up to the second loop.
An important by-product of these calculations is a systematic procedure
to evaluate the  Hikami boxes, see \FigsTwo{fig:Hikami}{fig:Hikami6},
which is based on a fundamental conservation law\cite{Hastings_Inequivalence_1994}:
electroneutrality of the density response function.
Our main analytical results for the quantum corrections to the polarizability
are presented in \EqsFromTo{eq:ChiRPASF}{eq:defF} with \EqsThree{eq:GCE1}{eq:GCE2}{eq:CE1}.

We have demonstrated that, in the experimentally relevant parameter range,
the difference between the statistical ensembles is unimportant and one
can fit the measurements by using the leading term of the perturbation theory.
The authors of
\RefsTwo{Deblock_Experiment_2000}{Deblock_Experiment_2002} have tried to
find 0D dephasing with the help of an empirical fitting formula. By using
the more rigorous and reliable \Eq{eq:DeltaFGCEFlux}, we have confirmed
that 0D dephasing might have manifested itself in the $T$-dependence of
magneto-oscillations at $ T \lesssim \ETh $. Unfortunately, the $T$-range
of interest here is rather narrow, and only few experimental data points are available there.
Therefore, we are unable to claim conclusively that 0D dephasing has been
observed in the experiments.
However, we can straightforwardly suggest several experiments
which might yield conclusive evidence of 0D dephasing:
First, one can repeat the measurement of \RefsTwo{Deblock_Experiment_2000}{Deblock_Experiment_2002},
but with a larger number of data points
around the crossover temperature $\Tcross$, see \Fig{fig:TempFit},
while simultaneously improving the measurements precision.
Since the theory predicts a drastic increase in slope of the
$\phi_0/2$-oscillations at the crossover (from $T^{-3/2}$ to $T^{-3}$),
even such measurements should be able to reliably confirm the existence
of 0D dephasing, thereby uncovering the role of the Pauli blocking at low $T$.
Second, it is highly desirable to extend the $T$-range where the
crossover to 0D dephasing is expected to appear, which can be achieved
by decreasing $\wo$ and/or increasing $g$. However, a very large
conductance and ultra-small frequencies are nevertheless undesirable,
because in these limits the quantum corrections to the polarizability are reduced. 
Thus, improving the precision of the measurement is needed anyway.
Besides, fitting with the help of the leading perturbative result fails
at very small frequencies, see \Fig{fig:Validity}. This difficulty can be
overcome by taking into account our two-loop results and/or using an
interpolation to the $\gamma \to 0$ limit from the RMT+$\sigma$-model,
see \Fig{fig:Result}.

To summarize, we have shown that the quantum corrections to the polarizability
are an ideal candidate to study dephasing at low $T$ and the crossover
to 0D dephasing. We very much hope that our theoretical results will
stimulate new measurements in this direction.

\begin{acknowledgments}
We acknowledge illuminating discussions with C.~Texier, H.~Bouchiat,
G.~Montambaux, and V.~Kravtsov, and support from the DFG through SFB
TR-12 (O.~Ye.), DE 730/8-1 (M.~T.) and the Cluster of Excellence,
Nanosystems Initiative Munich. O.~Ye. and M.~T. acknowledge hospitality
of the ICTP (Trieste)  where part of the work for this paper was carried out.
\end{acknowledgments}

\end{document}